\pdfoutput=1
\documentclass[12pt]{article}
\usepackage{sbc-template}
\usepackage{graphicx,url}
\usepackage[latin1]{inputenc}
\usepackage{latexsym}
\usepackage{amstext}
\usepackage{fancyheadings}
\usepackage{subfigure}
\usepackage[latin1]{inputenc}
\usepackage[T1]{fontenc}
\usepackage{indentfirst}
\usepackage{amsmath}
\usepackage{amsbsy}
\usepackage{bm}
\usepackage{amssymb}
\usepackage{tikz}
\usepackage{tkz-tab}

\usepackage{epsf}
\usepackage[dvips]{epsfig}

\newcommand{\eat}[1]{}

\usepackage{comment}

\usetikzlibrary{automata,arrows,shapes,calc,positioning}

\title{Modeling and Analysis of Cloud Signaling Services}

\author{Eduardo M. Hargreaves\inst{1 }, Paulo H. de Aguiar Rodrigues\inst{1}, Daniel S. Menasch\'e\inst{1} }
\address{Departamento de Ci\^encia da Computa\c{c}\~ao\\
  Universidade Federal do Rio de Janeiro (UFRJ),    Rio de Janeiro, Brasil
\email{eduardo.hargreaves@ppgi.ufrj.br, aguiar@ufrj.br, sadoc@dcc.ufrj.br}}

\begin{document} 
\maketitle

\begin{abstract}
 Networks connecting distributed cloud services through multiple data centers are called cloud networks. 
 These types of networks play a crucial role in cloud computing and a holistic performance evaluation is essential before planning a 
 converged network-cloud environment. We analyze a specific case where some resources  can be centralized 
in one datacenter or distributed among multiple data centers.  The economy of scale in centralizing resources in a single 
pool of resources can be overcome by an increase in communication costs. 
We propose an analytical model to evaluate tradeoffs in terms of application requirements, usage patterns, number of resources and communication costs.
We numerically evaluate the proposed model in a case study inspired by the oil and gas industry, 
indicating how to cope with the tradeoff between statistical multiplexing advantages of centralization and the corresponding increase in communication infrastructure costs.
\end{abstract}

\section{Introduction}
\label{sec:Intro}

Cloud computing is a large-scale distributed computing paradigm that is
driven by economies of scale, in which a pool of computing resources (e.g., networks, servers, storage, applications, and services) 
are provisioned, delivered and released on demand to users over a network \cite{foster2008cloud}. 
To the user, the available capabilities often appear to be unlimited and can
be elastically provisioned in any quantity at any time \cite{mell2011nist}. 

To improve cloud performance and resiliency, the current trend is to deploy the services across multiple and geographically distant sites
\cite{CloudNetworks}.
In this context, cloud networks (networks connecting cloud services hosted in multiple data centers) play a crucial role in cloud computing 
and are an indispensable ingredient
for high performance cloud computing \cite{duan2012survey}.

This work is motivated by a real-world project of deployment of an Oil \& Gas \emph{(O\&G)} application in multiple sites. Because  \emph{O\&G} possesses a high cost per license, buy one license per user can be very expensive. Because a user can run multiple applications, like e-mails clients or spreadsheets softwares, the single user licensing aproach results in a waste of resources due to license underutilization. One solution often used in corporate environments for cost minimization is share a license between multiple users with the adoption of the floating licensing approach \cite{hamadani1998automatic}. 
In this approach, a limited number of licenses is stored in a pool shared among a large  
number of users.
 When a user wishes to run the application, a license is requested to a license server.
The task of a software license server is to determine and control the number of active replicas of the software based on the 
license entitlements that an organization owns or the system capabilities. If a license is available, the license server allows the application to run. 
When the user finishes the application, the license is reclaimed by the license server 
and made available to other users. To avoid user's misbehavior, the application periodically sends license renewal requests to the license server.
If the client machine does not receive an answer, a timeout is detected and the application is closed.

The sharing of scarce resources (licenses) among users taking advantage of the probabilistic pattern of user's access produces  \emph{statistical multiplexing gains}. The sharing  increases the utilization of the licences and reduces the number needed to satisfy users requirements. This gain can be viewed as a consequence of the law of large numbers and increases as the number os users sharing a resourse increases. It is important to notice that there are more users than licenses,    therefore it is possible that a user cannot get an available license. In this case he waits or he is blocked. Systems of the first type are called \emph{waiting systems} while the second type systems are called \emph{loss-systems} \cite{teletraffic}. We modelled the licensing system as a \emph{loss-system}.  The probability of not get a license is the \emph{blocking probability}. Because blocking is inconvenient to users, the blocking probablity is a Service Level Agreement (SLA) parameter.

\begin{figure}[h]
\begin{center}
    \includegraphics[width=0.66\textwidth]{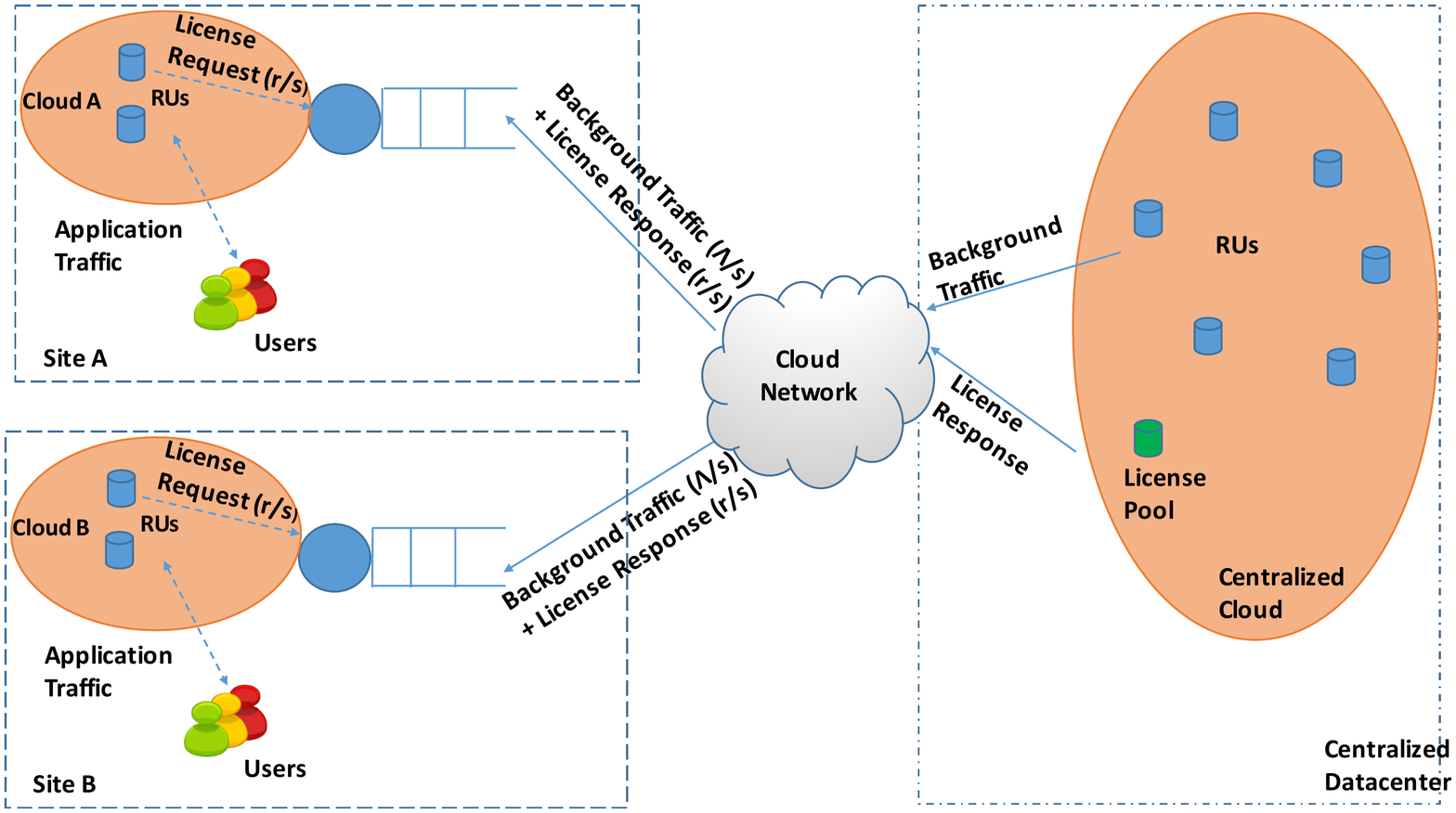}
\end{center}
\caption{Centralized  servers (single pool)}
\label{CentralizedLicenseServers}
\end{figure}

\begin{figure}[h]
\begin{center}
    \includegraphics[width=0.66\textwidth]{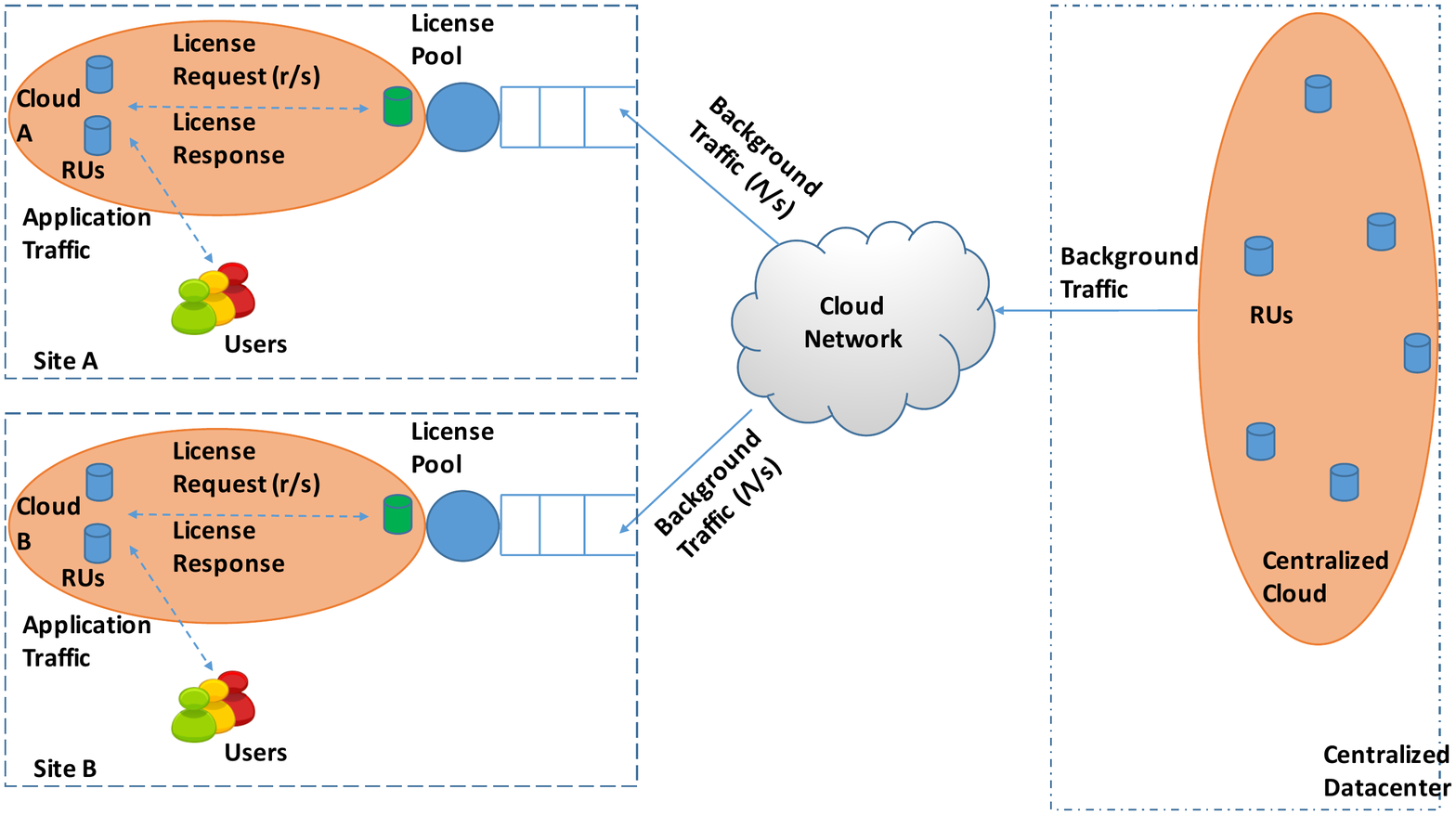}
\end{center}
\caption{Distributed servers (multiple pools)}
\label{RemoteLicenseServers}
\end{figure}

A cloud is as a set of one or more \emph{resource units (RUs)} running services that are requested by users. Application server and licence server are examples of servers running in RUs.
The set of \emph{O\&G} applications is so image intensive and demands so much 
  bandwidth  that  users  and RUs running the application servers have to be in the same local area network due to performance constraints.
   The cloud architecture can be a federated network of public or private clouds, hosting the software floating license cloud service, the \emph{O\&G} application itself and other corporative applications.  Licenses are consolidated remote to users in a single pool  (Figure \ref{CentralizedLicenseServers})
or near to users spread over multiple pools in multiple sites (Figure \ref{RemoteLicenseServers}).

The single pool and multiple pool architectures considered in this paper make use  of soft-state signaling solutions~\cite{ji2003comparison}.  In soft-state solutions, installed state at a remote site needs to be frequently refreshed by  clients running the  application.  Otherwise,  the state times out (and is removed).  
Soft-state solutions are quite common, being used in a variety of protocols as RSVP, SRM, SIP and IGMP, to name a few. In hard-state solutions, in contrast, state remains installed until explicitly removed.  Soft-state requires well-provisioned links for frequently updating states at remote sites through a WAN.

In the single pool case, the federated clouds are linked by the network cloud in 
such a way that background traffic
competes with the request/response signaling traffic. Due to congestion generated 
by the background traffic,  timeouts can occur and 
the application might be prematurely closed. 
In this situation, the cloud network capacity needs to be increased to reduce the premature 
timeout probability.   In the multi-pool case,   users access local (but typically more scarce) resources at distributed pools,  the connection is typically over over-provisioned local area networks (LANs) and timeouts due to congestion rarely occur.  
While the single pool explores the
statistical multiplexing gains due to resource sharing, such economy of scale gains might be percluded by an increase in communication costs. 
\begin{enumerate}
\item How to quantify  advantages and disadvantages of consolidating resources in a single 
pool?
\item What is required (in terms of resources and infrastructure upgrades) to satisfy the service level agreements? 
\end{enumerate}

While partially answering the questions above, our key contributions are the following:

\begin{enumerate}
\item \textbf{holistic analysis of license serving: } we propose an integrated framework for the assessment of benefits and costs of service infrastructures accounting for  software and network aspects. We apply our framework to the analysis of how to distribute resources in a cloud environment;
\item \textbf{analytical model: } we specialize the proposed framework to analyze the tradeoffs involved in resource distribution accounting for   
gains due to statistical multiplexing  and costs due to congestion;
\item \textbf{case study: } we numerically investigate the proposed model using a case inspired by a real-world Oil \& Gas setup.
\end{enumerate}

The remainder of this paper is organized as follows. Related work is
 discussed in Section \ref{sec:trabalhos_relacionados}.
In Section \ref{sec:modelo} we present the proposed analytical model. 
Section \ref{sec:exemplos} contains the numerical 
examples, Section~\ref{sec:discu} presents further discussion about the model applicability   and Section \ref{sec:conclusions} concludes the paper.

\section{Related Work}
\label{sec:trabalhos_relacionados}

Since networking has a strong impact on end-to-end cloud service, a holistic performance evaluation is essential before planning a 
converged network-cloud environment. Performance evaluation on the sufficient number of computational resources necessary to meet a desired 
Service Level Agreement (SLA) in a cost-effective way has attracted extensive research interest \cite{brandwacht2013models,khazaei2012performance}.
\cite{duan2014holistic} has proposed a modeling and analysis approach by exploiting network
calculus theory to define a general profile that can characterize service capability of either a network or Cloud service.

Our work is closely related to  the facility location problem in operations research~\cite{facloc}.   In facility location problems, the aim is optimally place facilities so as to minimize transportation costs. While we also target minimization of service costs, we consider the interplay between those and statistical gains due to multiplexing, timeout probabilities due to congestion and blocking probability due to resource exhaustion. 

\cite{khazaei2012performance} modeled the blocking probability of a cloud with a large number of Resource Units (RUs) and general service times given by 
a $M/G/m/m+r$ queuing system with single task arrivals and a task buffer of finite capacity.
\cite{brandwacht2013models}
evaluated finite population and heterogeneous resource requests using the blocking probability as an SLA performance measure to dimension clouds. 
They showed that infinite source models may lead to an overestimation of the number of RUs.

Although the papers above are related to ours, none of them considered the tradeoff between gains due to statistical multiplexing  and costs to support remote traffic.  
While~\cite{khazaei2012performance} and ~\cite{brandwacht2013models}
  considered  blocking of RUs,~\cite{duan2014holistic} considered  network-cloud service capabilities.  
To the best of our knowledge, this paper is the first to bridge the two aspects in an integrated manner. 

\section{Model and Notation}
\label{sec:modelo}

We begin by describing the system of interest in this paper.  A cloud is a pool of one or more RUs that can be requested by users. These units can be virtual machines (VM) 
or CPUs. We assume that there are RUs acting as signaling (control) servers and application servers. Each signaling  server controls one pool of 
application states. To simplify presentation, we assume that each running instance of the application is associated to an application state.

\begin{table}[h!]
\center
\begin{tabular}{ll}
\hline 
\hline
variable & description \\
\hline
$N$ & number of sites \\
$S$ & population size \\
$S_i$ & population at site $i$ \\
$L$ & number of supported application states (resources) \\
$L_i$ & number of supported application states (resources) at pool $i$ \\
$\lambda$ &  request arrival rate per user (requests/hour)\\
$1/\mu$ & average duration of a session (hours)\\
$\rho$ & $\lambda/\mu$\\
$C_0$ & initial circuit capacity (Mbps)\\
$C'$ & additional circuit capacity (Mbps)\\
$C$ & circuit capacity (Mbps) $(C=C'+C_0)$ \\ 
$1/M$ & packet mean size ($\textrm{bits}^{-1}$)\\
$\Lambda$ & background packet arrival rate (pkts/s)\\
$r$ &rate at which application states are checked $(\textrm{s}^{-1})$\\
$\tau$ & timeout detection threshold (s) \\
\hline
\hline
metric & description \\
\hline
$b_c$ & blocking probability in the centralized setup \\
$b_d$ & blocking probability in the distributed setup \\
$p$ & timeout probability in the distributed setup \\
$s_c $ & success probability in the centralized setup \\
$s_d $ & success probability in the distributed setup \\ 
$s $ & required success probability defined in SLA \\
\hline
\end{tabular}
\caption{Table of notation}
\label{Notation}
\end{table}

We consider a population of $S$  users divided into $N$ sites (or pools in the distributed case).  
Each site $i$ comprises $S_i$ users.  
Each user initiates new application instances at rate $\lambda$. 
If the  signaling server is able  to allocate space to store state information associated to the new instance, the requester  accesses the VM until the service is completed. When the user finishes its session, the space used to store state information is reclaimed by the license server and made available to other users. Each session duration is exponentially distributed with mean $1/\mu$ hours. Let $\rho=\lambda/\mu$. Table \ref{Notation} summarizes the notation used throughout the rest of the paper.

Based on observed real-world signaling packets, application periodically send state renewal requests to the signaling servers at fixed rate of $r$ attempts/s and detects a timeout if it does not receive an answer  after $\tau$ seconds.
In the single  pool architecture 
requests traverse a circuit with capacity $C$ competing with background traffic.  Background packets arrive at rate $\Lambda$ pkts/s and each packet requires exponentially distributed service 
with mean $1/M$.

\subsection{Blocking probability}

% \begin{equation}
% P(bloqueio)=\sum^{M}_{L+1} \binom{M}{L+1}p^{L+1}(1-p)^{M-(L+1)}
% \end{equation}

\begin{figure}
\begin{tikzpicture}[->,>=stealth',shorten >=1pt,node distance=2.25cm,
	    thick,main node/.style={circle,draw,font=\sffamily\scriptsize\bfseries}]
	    
	    \node[main node] (0) {$0$};
	    \node[main node] (1) [right of=0] {$1$};
	    \node         (d1)  [right of = 1]   {$\cdots$};
	    \node[main node] (i) [right of=d1] {$i$}; 
	    \node (d2) [right of=i] {$\cdots$};
	    \node[main node,text width=0.55cm] (n-1) [right of=d2] {\tiny$L-1$};
	    \node[main node] (n) [right of=n-1] {$L$};

	      \path[style={font=\sffamily\small}]
		(0) edge[bend left,below] node [above]{\scriptsize$S\lambda$} (1)
		(1) edge[bend left,below] node [above]{\scriptsize$(S-1)\lambda$} (d1)
		    edge[bend left,above] node [below] {\scriptsize$\mu$} (0)
		(d1)  edge[bend left,above]node [below] {\scriptsize$2\mu$} (1)
		    edge[bend left,below] node [above]{} (i)
		(i) edge[bend left,below] node [above]{\scriptsize$(S-i)\lambda$} (d2) 
		     edge[bend left,above]node [below] {\scriptsize$i\mu$} (d1)	
		(d2)  edge[bend left,above]node [below] {} (i)
		    edge[bend left,below] node [above]{} (n-1)
		(n-1)  edge[bend left,below] node [above]{\scriptsize$(S-L+1)\lambda$} (n) 
		     edge[bend left,above]node [below] {\scriptsize$(L-1)\mu$} (d2)	   
		(n)  edge[bend left,above]node [below] {\scriptsize$L\mu$} (n-1);	    
\end{tikzpicture}
\caption{State transition diagram for the finite source model}
\label{fig:engset}
\end{figure}
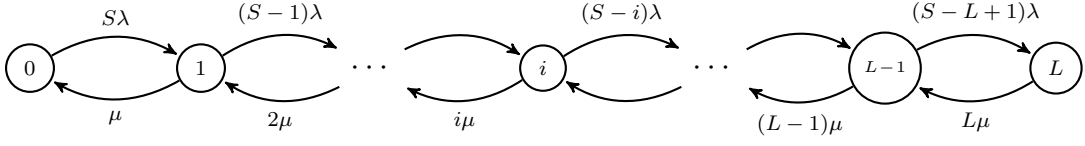

To characterize the blocking probability, we use a finite source queueing model with homogeneous service 
distribution and no waiting time, 
as given by the well-known \textit{Engset formula} \cite{teletraffic}. The aggregated arrival 
rate of renewal requests is proportional to the number of idle users and the aggregate service rate is proportional to active users to whom access was granted. 
The state transition diagram of the finite source model is illustrated in Figure \ref{fig:engset}.

The states of the application replicas   can be consolidated in a single pool or distributed across multiple pools. 
Let $b(L,S)$ be the blocking probability for a population of size $S$ associated to a  signaling server that has capacity to handle  $L$ application states.  
\begin{equation}
b(L,S)=\frac{\binom{S-1}{L}(\frac{\lambda}{\mu})^L}{\sum^{L}_{i=0}\binom{S-1}{i}(\frac{\lambda}{\mu})^i}
\label{cent_blocking}
\end{equation}

The numerical calculation of \eqref{cent_blocking} leads to numerical problems for large values of $S$ and $L$.
So, to compute $b$, we used the following numerically stable recursive formula~\cite{teletraffic}: 
 
 \begin{equation}
b(j,S)=\frac{\rho(S-j+1)b(j-1,S)}{j+\rho(S-j+1)b(j-1,S)}, \textrm{ where } b(0,S)=1 \textrm{ and } 1\le j \le L
\label{cent_blocking_numerical}
\end{equation}

When referring to the centralized setup,  we might denote $b(L,S)$ simply by $b_c$.
In the centralized case, $b_c$ is given by:
\begin{equation}
b_c=b(L,S)
\end{equation}

In the distributed scenario, there is one signaling server at each pool.  Users compete for access to the signaling server of their corresponding pools and each pool $i$ has its associated blocking probability $b(L_i,S_i)$. Let $b_d$ be the blocking probability in the distributed case, given as the weighted sum of the blocking probabilities at each pool.  
So, the blocking probability in the distributed case is given by:
\begin{equation}
b_d =  \sum_{i=1}^N  \frac{S_i}{S} b(L_i,S_i) 
\label{mult_blocking}
\end{equation}
where
\begin{equation}
S=\sum^N_{i=1}S_i \,\,\, , \,\,\, L=\sum^N_{i=1}L_i 
\end{equation}

\subsection{Timeout Due To Congestion}

In this section we assume that each client periodically sends state renewal requests 
to the signaling server, at a constant rate of  $r$ requests per second.
Each renewal request is also referred to as a \emph{probe}.  
If the client does not receive an answer in $\tau$ seconds ($\tau < 1/r$), a timeout is detected 
and the application is closed. 
Note that due to network congestion, the chances of occurring a timeout if 
the signaling server is
centralized is higher than if the signaling  server is closer to end-users.

%In the centralized scenario, due to congestion, the waiting time in queue
% can be greater than $\tau$ seconds. The question addressed here
% is the probability of timeout due to congestion.}

Let $D_i$ be the duration of the $i$-th subinterval in which the state is checked.
Let $\tilde{N}$ be the number of times that probes are sent to the server until a timeout is 
generated, without counting the last probe ($\tilde{N} \ge 0$).
Let $q$ be the probability that a probe successfully yields a state renewal. Then,
   $\tilde{N} \sim Geometric(1-q)$, 
\begin{equation}
P(\tilde{N}=n)=q^n (1-q) \,\,,\,\, n \ge 0
\end{equation}
Let $D$ be the time until a timeout, where $D$ is given as a function of $D_i$ and $N$ 
as follows,
\begin{equation}
  D = D_0 + D_1 + D_2 + \ldots + D_{\tilde{N}}  + D_{\tilde{N}+1} \\
  \label{D}  
  \end{equation}
  where $D_0=0$.   
In the remainder of this section we assume that 
$D_i\sim Constant(\frac{1}{r})$ for $i=1, 2, 3, \ldots$ but 
the argument presented below easily generalizes for the case where $D_i$ has general distribution, 
and  $D_1, D_2, \ldots$ are independent and identically distributed with Laplace transform given by $e^{(-s/r)}$. 

Conditioning on $\tilde{N}$, 
the Laplace transform of $D$ can be derived as:
\begin{equation}
   D^{\star}(s) =  \tilde{N}(z)\bigg|_{z=e^{-(\frac{s}{r})}} D_{\tilde{N}+1}^{\star}(s) \\
  \label{D_star}
  \end{equation} 
where $\tilde{N}(z)$ is the z transform of $\tilde{N}$,
   \begin{eqnarray} \label{eqnz}
     \tilde{N}(z) &=&  \frac{1-q}{1- z q} 
  \end{eqnarray}
and
\begin{equation} \label{eqdi}
    D_{\tilde{N}+1}^{\star}(s) = D_i^{\star}(s) = e^{-s/r}, \qquad i=1, 2, 3, \ldots
\end{equation}
We substitute~\eqref{eqnz} and ~\eqref{eqdi}  into~\eqref{D_star} to obtain
\begin{eqnarray}
    D^{\star}(s) &=& e^{-{s}/{r}}\bigg( \frac{1-q  }{1- e^{-{s}/{r}}q }\bigg) 
%     D(k) &=& p^{k-\frac{1}{r}}(1-p)
\label{D_star_fim}
\end{eqnarray}

Next, we characterize the network congestion which will cause the timeouts. This characterization is the first step in deriving a formula for the 
timeout in terms of $C$ in such a way that tradeoffs can be evaluated in terms of $L$ and $C$.

  Let $Y$ be the delay experienced by a probing packet that traverses a link with capacity
   $C$  subject to  exogenous (background) traffic with arrival rate $\Lambda$ and size exponentially distributed with mean $1/M$.  
  The CDF of $Y$ is given by  $F_Y(y)=P(Y < y)$.  Then, the probability that a 
  state renewal request
  succeeds is given by $q=P(Y  < \tau)$. 
  In the remainder of this paper,
   we assume that the delay is characterized by an 
   M/M/1 queue, and that the overhead caused by 
 state renewal requests into the link is negligible compared to the exogenous traffic. 
 Then, it follows from~\cite[equation (5.119)]{kleinrock} that
 \begin{equation} 
q=1-e^{-( MC-\Lambda)\tau}
\label{MM1_waiting_time}
\end{equation}  

The \emph{target session duration}  is the desired duration of the session as determined
by the user requirements.   
Let the target session duration time be exponentially distributed with mean $1/\mu$,
 and let $L$ be the random variable that characterizes
the target session duration, $L \sim Exponential(\mu)$.  The actual session duration 
might be smaller than the target session duration if a timeout occurs.  

Let $p$ be the probability of timeout. A timeout occurs if the target session duration is greater than the time until a renewal
request fails.  Therefore, $p$ is given by: 
  \begin{equation}
  p = P(L>D)= D^{\star}(s) \Big|_{s=\mu} = D^{\star}(\mu)
  \label{P_timeout1}
  \end{equation}
 where $D^{\star}(\mu)$ is the Laplace transform of $D$  given by \eqref{D_star}, evaluated
 at the point $s=\mu$.
 
 Our goal is to obtain a simple expression for $p$ as a function of $C$. 
To this aim, we substitute~\eqref{MM1_waiting_time} into \eqref{D_star_fim}  and make use
of  \eqref{P_timeout1} to obtain $p$,
   \begin{eqnarray}
    p &=&  e^{-\mu/r}\left( \frac{e^{-( MC-\Lambda)\tau}  }
    {1- e^{-{\mu}/{r}}\left(1- e^{-( MC-\Lambda)\tau}\right)}\right) 
  \label{P_timeout_final}
  \end{eqnarray}  
  Note that if $q=0$ then $p=e^{-\mu/r}$. 
  This is an expected result, as $q=0$ means that the state renewal will fail and timeout will only occur if the target duration is greater than 
  $1/r$, the interval between renewal requests.
  The probability that
  the application cannot be  started  at first place is captured through $b_c$.
  
%    \begin{eqnarray}
%     D(z) &=&  N(z) D_i(z)\\
%     N(z) &=& \frac{F(t)z  }{1- z F(t)} \\
%     D_i(z) &=& z^{(\frac{1}{r})}\\
%     D(z) &=& z^{\frac{1}{r}}\bigg( \frac{F(t)z  }{1- z F(t)}\bigg) \\
%     z^{\frac{1}{r}}\bigg( \frac{F(t)z  }{1- z F(t)}\bigg) &\Longleftrightarrow& N(k-\frac{1}{r})\\    
%     D(k) &=& N(k-\frac{1}{r})= F(t)^{k-1-\frac{1}{r}}(1-F(t))\\
%   \end{eqnarray}
%   
  
  After some algebraic manipulation, it is possible to derive an explicit formula for the required link capacity $C$ as a  function of the targeted timeout probability   $p$, 
  the application characteristics ($\tau$ and $1/r$), usage 
  patterns ($\mu$) and background traffic characteristics ($\Lambda$ and $M$):
  \begin{equation}
  C = \frac{1}{M}\bigg(\Lambda +\frac{1}{\tau}\ln\bigg(\frac{p(e^{\frac{\mu}{r}}-1)}{p-1}  \bigg)\bigg)
  \label{Capacity}
  \end{equation}

% \begin{equation}
% P(bloqueio)=\frac{\binom{M_1}{L}(\frac{\lambda}{\mu})^{L_1}}{\sum^{L_1}_{i=0}\binom{L_1}{i}(\frac{\lambda}{\mu})^i}+ \frac{\binom{M_2}{L}(\frac{\lambda}{\mu})^{L_2}}{\sum^{L_2}_{i=0}\binom{L_2}{i}(\frac{\lambda}{\mu})^i}
% \end{equation}
\subsection{Success Probability}

We consider the probability of success $s$ the main SLA parameter. The success is defined by being granted access to the application  and being successful in all
attempts of state renewal.

Let $s_c$  be the probability of success in the centralized setup. This probability is the product of the probability of not being blocked (1-$b_c$)  
times the probability that all attempts succeeded (1-$p$), with $b_c$ and $p$ given by \eqref{cent_blocking} and \eqref{P_timeout_final}.  
\begin{equation}
s_c=(1-b_c) (1-p) 
\label{psuccess_cent}
\end{equation}

Let $s_d$ be the probability of success in the  distributed scenario. As the application and the signaling server are in the same cloud
$p$ is zero. On the other hand, the signaling servers at each pool have less resources than the central server, which  might increase the probability of blocking.
Then, $s_d$ is given by: 
\begin{equation}
s_d=1-b_d
\label{psuccess_distributed}
\end{equation}

In light of equations \eqref{psuccess_cent} and \eqref{psuccess_distributed}, we are ready to quantify the tradeoff mentioned in the beginning of this paper.  The centralized setup 
is associated to a smaller blocking probability as resources are multiplexed in a single pool. Nonetheless, users incur a  
timeout probability due to network congestion.  This tradeoff motivates an optimization problem, where the network designer is faced with a decision 
between centralizing the  pool of resources or distributing resources across multiple pools.

\subsection{Optimization Problem}
\label{sub:optimization}

The optimization problem consists of minimizing costs for a given success probability defined in an SLA. Formulas 
\eqref{cent_blocking}, \eqref{mult_blocking} and \eqref{P_timeout_final} are consolidated in equations 
\eqref{psuccess_cent} and \eqref{psuccess_distributed}. These equations
give the success probability in terms of application usage patterns, number of users, licenses and pools, renewal attempts characteristic, capacity
and traffic of the network cloud. The communication and resource costs  determine the  merits of the centralized and distributed scenarios.

Next, we present the resource allocation problem in the centralized scenario:
\begin{eqnarray}
\textrm{minimize} &:& \,\,\,\,\, c=\alpha L+\beta nC' \label{cost1} \\
\textrm{subject to}  &:&  \,\,\,\,\,s_c \ge  s\\
\textrm{constraint on variables}  &:&  \,\,\,\,\,L \ge 0, \,\, C' \ge 0
\label{optimization_problem_sc}
\end{eqnarray}

The corresponding distributed resource allocation problem is: 
\begin{eqnarray}
\textrm{minimize} &:& \,\,\,\,\, c=\alpha L+\beta nC'\\
\textrm{subject to}  &:&  \,\,\,\,\,s_d \ge  s\\
\textrm{constraint on variables}  &:&  \,\,\,\,\,L \ge 0, \,\, C' = 0
\label{optimization_problem_sd}
\end{eqnarray}
where $c$ is the cost in the centralized or distributed scenario,
$\alpha$ is the cost per maintained state (resource), $\beta$ is the cost per Mbps and $n$
is the number of sites that need to be upgraded in terms of capacity.  We consider a link with initial capacity $C_0$.   Let 
$C'$ be the marginal capacity added to the link.  Let $C=C_0+C'$ be the total link capacity (also referred to simply as \emph{link capacity}).  Note that the   timeout probability $p$ is a function of the total capacity $C$. 
The probability of success is given by \eqref{psuccess_cent} and \eqref{psuccess_distributed}  in the centralized and distributed scenarios, respectively. 
The scenario with the minimum cost between the centralized resource allocation problem and distributed resource allocation is the scenario with the total minimum cost.

Let $c_c^{\star}$ and $c_d^{\star}$ be the minimum costs achieved through the centralized and distributed setups, respectively.  
Then, the network designer chooses the centralized or the distributed setups so as to minimize the minimum feasible cost $c^{\star}$:
\begin{eqnarray}
c^{\star}= \min (c_c^{\star} , c_d^{\star}) 
\end{eqnarray}

In order to find the optimal solution, 
we used an \textit{interior point method} in which the derivatives
are approximated by a solver as implemented in  Matlab\textsuperscript{\textregistered} convex optimization toolbox.

\begin{comment}
\section{Private cloud remote to users}

In this scenario all servers are centralized in a private cloud located in a centralized datacenter. The question here is to evaluate the tradeoff of increasing capacity versus the economy of 
reducing both the quantity of servers and number os licenses.

\begin{figure}[h]
\begin{center}
    \includegraphics[width=0.6\textwidth, angle =270 ]{General_centralized.eps}
\end{center}
\caption{Private Cloud with Centralized Servers}
\label{P_timeout}
\end{figure}
\end{comment}

\section{Numerical Examples}
\label{sec:exemplos}
In this section we numerically investigate 
the proposed model.
Our goals are to a) numerically illustrate the tradeoffs between blocking probability and timeout probability and b) indicate the applicability of the  
optimization problem proposed, quantifying the advantages and disadvantages of central and distributed pools of resources.  Our examples in  this section 
are motivated by the  previously mentioned  Oil \& Gas application. In this application, signaling servers associate one license to each  running instance of an application.  Therefore, in what follows we refer to \emph{licenses} and  \emph{application state resources} interchangeably.

\begin{figure}
\begin{center}
  \begin{tabular}{l r}
   \subfigure[Blocking Probability]{
    \resizebox{0.48\textwidth}{!}{ 
	    \includegraphics[width=0.95\textwidth]{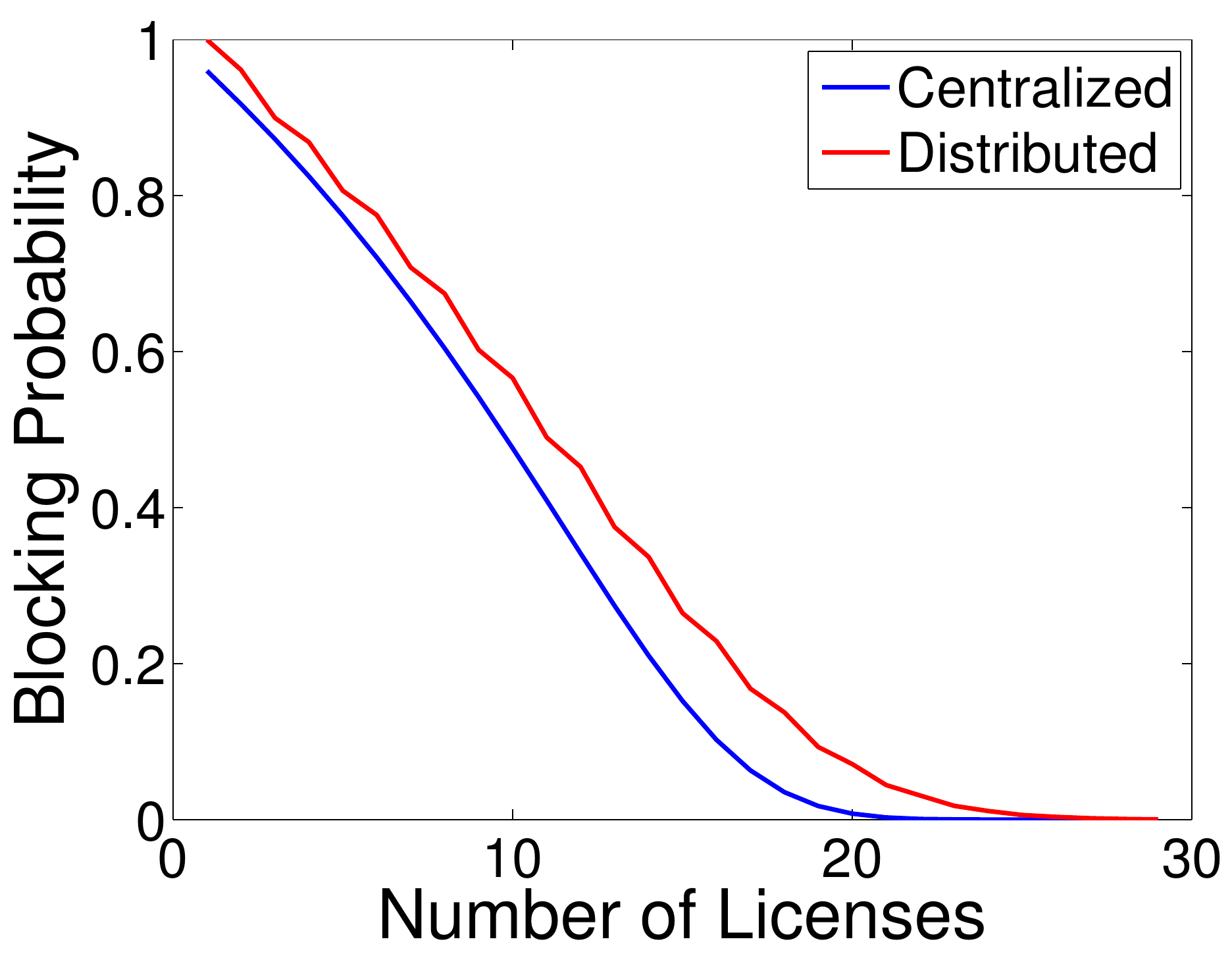}
	  }	  
	}  
  \subfigure[log(Blocking Probability)]{
    \resizebox{0.48\textwidth}{!}{ 
	  \includegraphics[width=0.95\textwidth]{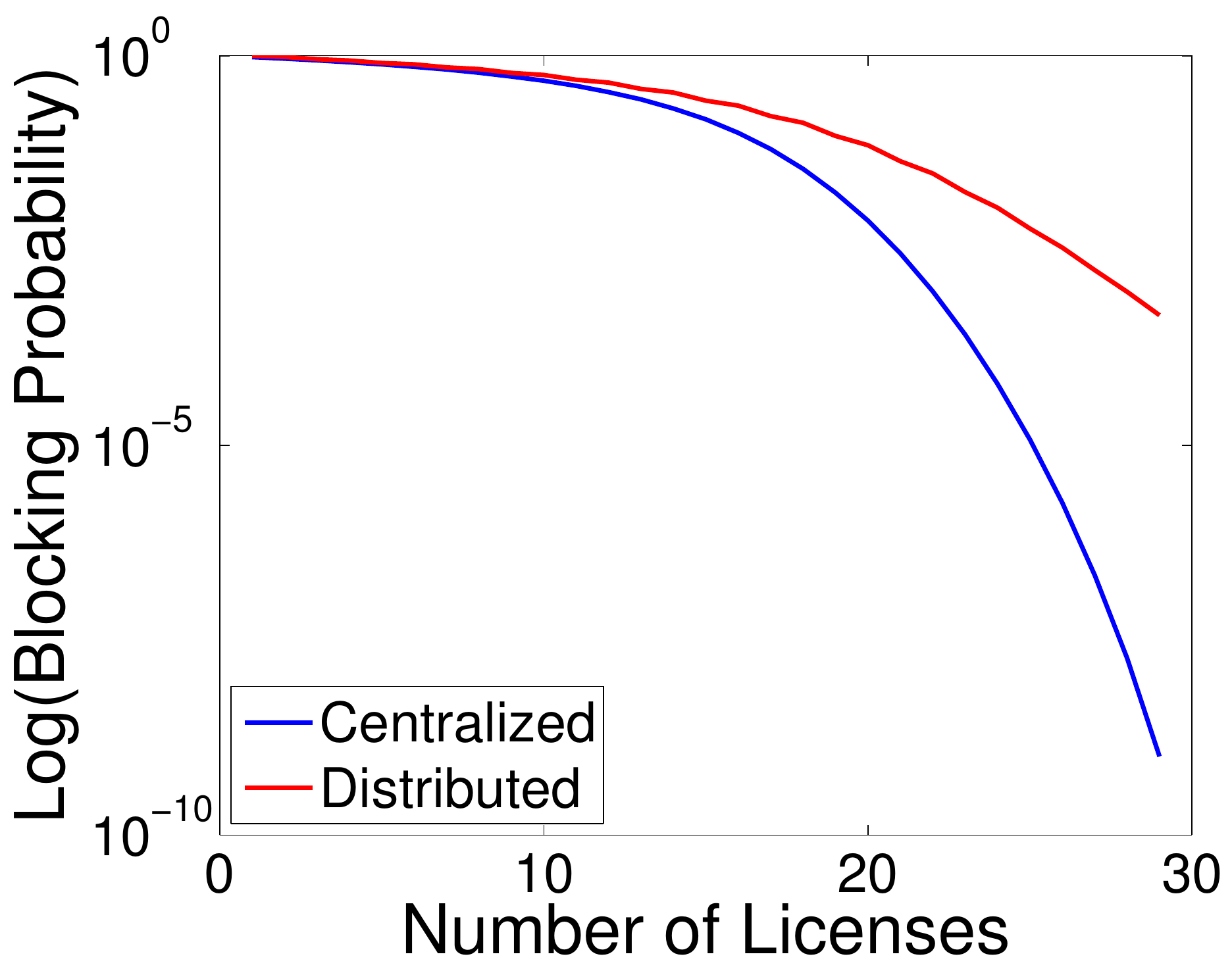}
	  }
	}  
\end{tabular}
\end{center}
\vspace{-0.2in}
\caption{Blocking probability, $\frac{1}{\mu} = 8$  hours, $S = 30$  users, $\rho=0.8$, blue=centralized, red=distributed versus number of 
licenses}
\label{fig:P_bloq}
\vspace{-0.2in}
\end{figure}

%\subsection{Centralized pool of licenses, typical example}
The first analysis is a comparison between the blocking probability in the centralized and in the distributed case as shown in Figure  \ref{fig:P_bloq}(a).
In terms of blocking probability, the centralized architecture is always better. The logarithm of the blocking 
probability,  plotted in Figure \ref{fig:P_bloq}(b), shows that the advantage of the centralized architecture increases as the number of available licenses increases.  We refer to the reduction in blocking probability due to centralization as \emph{licensing statistical multiplexing gains}.

\begin{figure}[h]
\begin{center}
\begin{tabular}{cc}
    \includegraphics[width=0.48\textwidth]{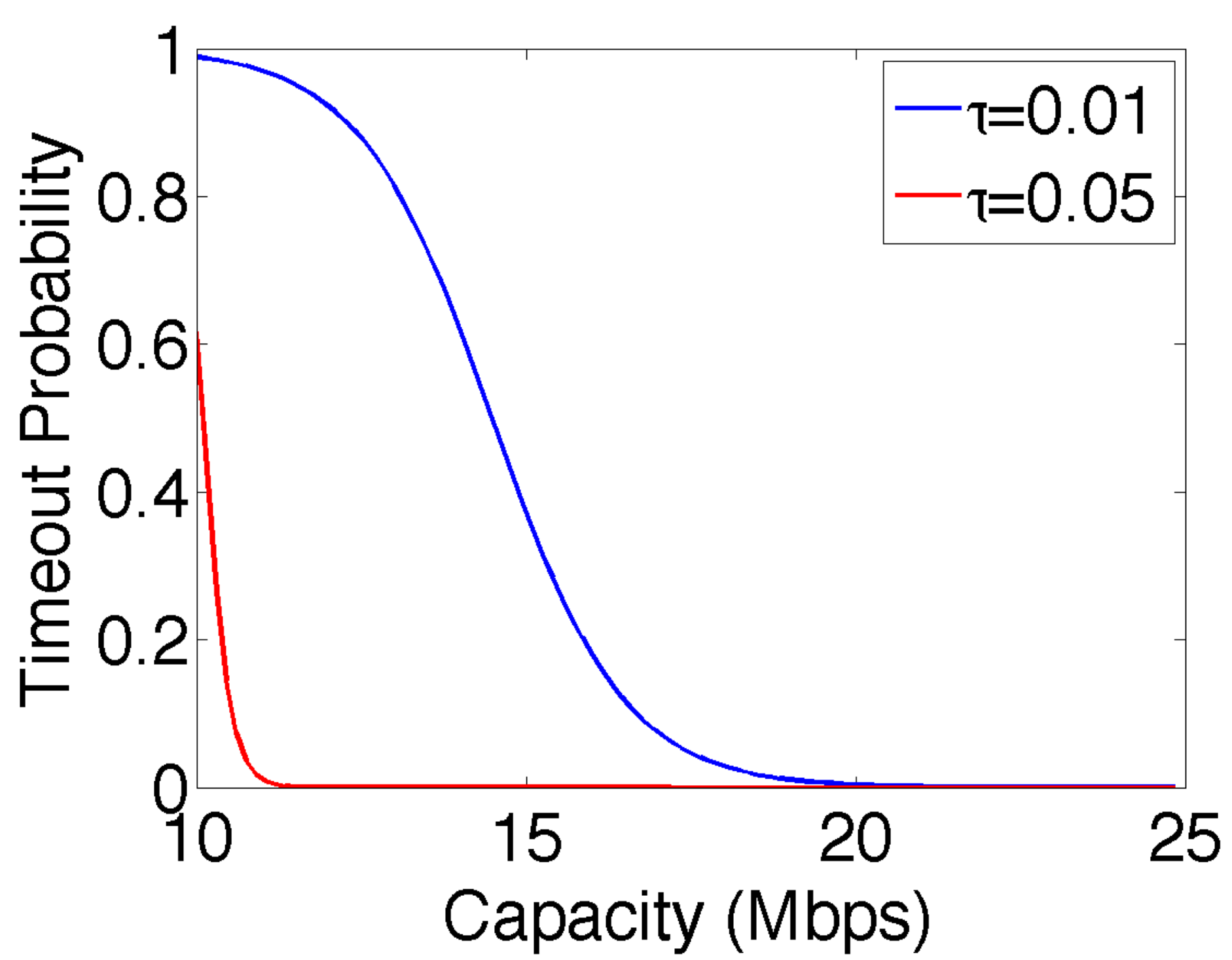}
& 
    \includegraphics[width=0.48\textwidth]{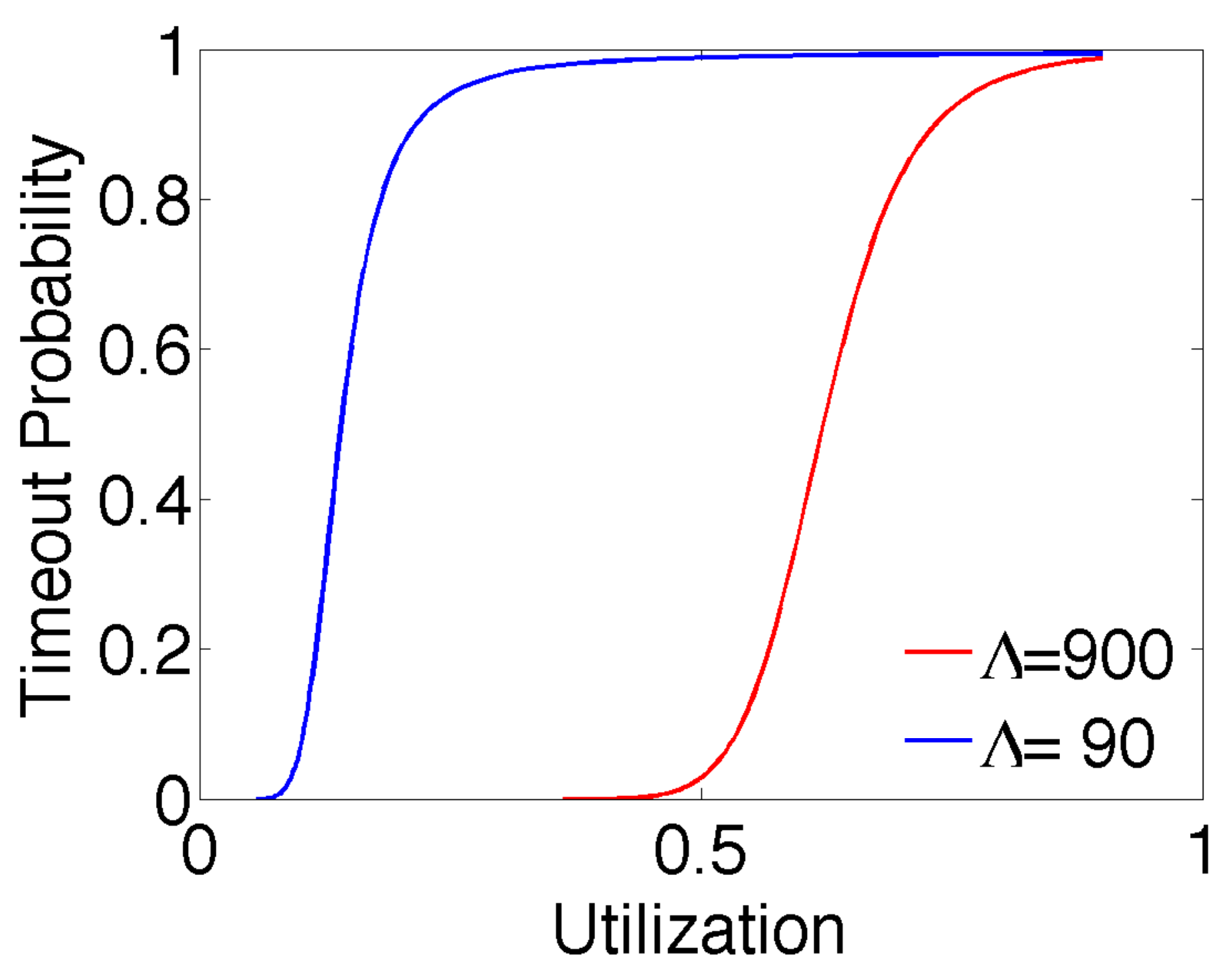} \\
    (a) & (b)
    \end{tabular}
\end{center}
\caption{Timeout probability due to congestion ($p$) for different values of (a) $\tau$  (for fixed $\Lambda=900$ pkts/s) and (b) $\Lambda$ (for fixed $\tau=0.01$s). $\frac{1}{\mu} = 8$  hours, $\frac{1}{r} = 120$ s,  $\frac{1}{M}=1250 $ Bytes.   In figure (b):  
red line, $\Lambda = 900$ pkt/s and $C \in [10,25]$ Mbps; blue line, $\Lambda = 90$ pkt/s and $C \in [1,15]$ Mbps.   }
\label{fig:P_timeout_X_rho} \label{fig:P_timeout}
\end{figure}

Figure \ref{fig:P_timeout}(a) shows the timeout probability due to congestion $p$ as a function of the link capacity $C$ for  $\tau=0.01$ and $\tau=0.05$. 
As the  capacity  $C$ increases,  $p$ decreases.  When $C = \infty$ (over-provisioning) we have $p=0$.  Figure \ref{fig:P_timeout}(a) also shows the significant impact of the timeout detection threshold $\tau$. When $\tau=0.05$, a small increase in capacity can reduce $p$
to zero. When $\tau=0.01$ it is necessary to double the capacity (from 10 Mbps to 20 Mbps) to achieve the same result. In as $\tau$ increases, $p$ sharply decreases to zero for small values of $C$.  On the other hand, when $\tau \approx 0$  the centralized architecture is infeasible. Roughly speaking, if an application developer wishes to allow the application to be used in different sites with a centralized floating 
licensing approach and the network capacity is small, the parameter $\tau$ must  be relaxed.

Figure \ref{fig:P_timeout_X_rho}(b) shows how the timeout probability due to congestion ($p$)  varies as a function of the the utilization factor for different workloads and link 
capacity ranges.
We vary the link capacity in the range of [10,25] (red line) and [1,15] (blue line) adjusting the offered workload accordingly.  Given a target value for $p$, the higher capacity network 
(red line)   can operate at a higher utilization level than the lower capacity network (blue line). 
We refer to this increase in    supported utilization  due to increased capacity as \emph{networking statistical multiplexing gain}. One of its consequence is that it is better to have one higher-speed link
instead of having n-parallel lower-speed links to carry the same amount of traffic \cite{pioro2004routing}.

\begin{comment}

\begin{figure}[h]
\begin{center}
\end{center}
\caption{Timeout probability due to congestion for different values of $\tau$. $\frac{1}{\mu} = 8$  hours, $\frac{1}{r} = 120$ s, ,$\Lambda = 900$  pkt/s, 
$\frac{1}{M}=1250 $ Bytes, blue= $\tau = 0.01$ s, red=$\tau = 0.05$ s   }
\label{fig:P_timeout}
\end{figure}

\end{comment}

\begin{figure}[h!]
\hspace{-0.25in}
\bgroup
\setlength\tabcolsep{0pt}  \begin{tabular}{cc} \setlength\tabcolsep{0pt} 
      \includegraphics[width=0.55\textwidth]{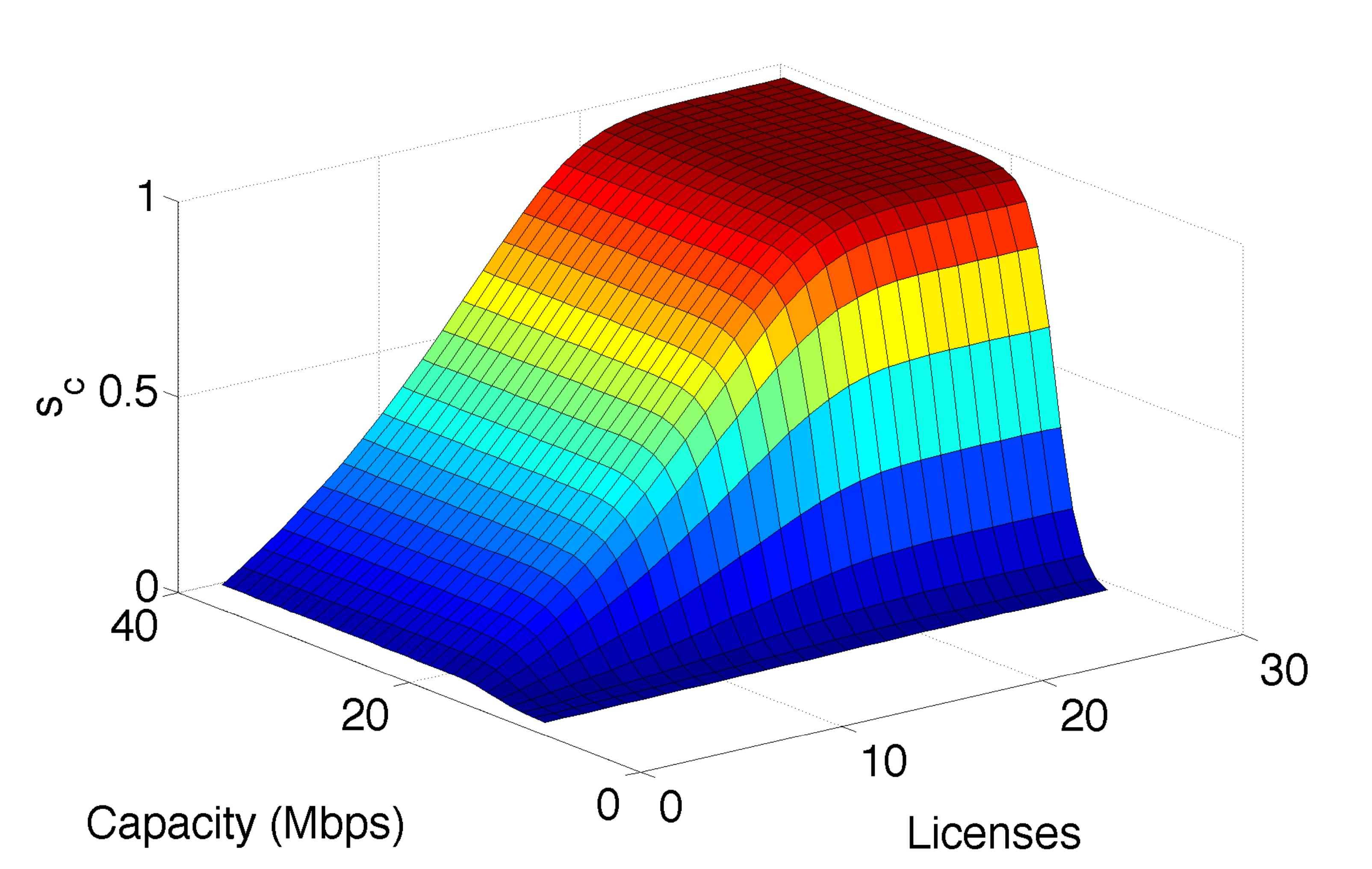} & \hspace{-0.18in}
	  \includegraphics[width=0.5\textwidth]{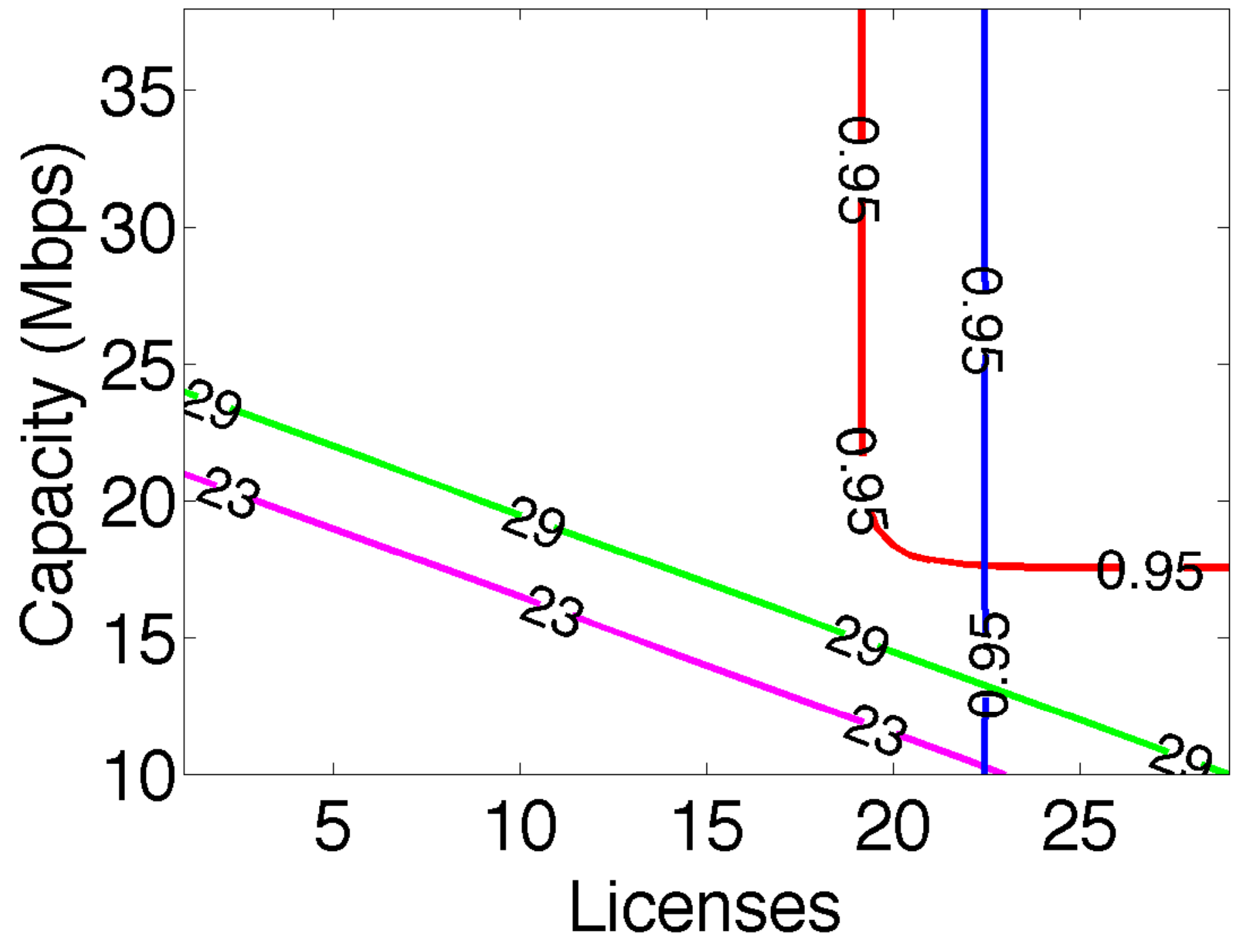} \\
	  (a) $s_c$ (centralized architecture) & (b) contour plots  \\
\end{tabular}
\egroup
\caption{Graphical solution of optimization problem: (a) $s_c$ as a function of capacity and number of licenses,  (b)  contour plots of $c_c=L+2C'$ (magenta and green lines), $s_c=0.95$ (red curve) and $s_d=0.95$ (blue curve).   ${1}/{\mu} = {1}/{\lambda}= 8$  hours, $S = 30$ users, 
$\frac{1}{\mu} = 8$  hours, ${1}/{r} = 120$ s, $\tau = 0.01$ s, ${1}/{M}=1250 $ Bytes, $\Lambda=900$, $C_0=10$ Mbps.   ``Capacity'' refers to $C=C_0+C'$.}
\label{fig:Optimizations1}
\end{figure}

Figure \ref{fig:Optimizations1}(a) shows the centralized probability of success $s_c$ as a function of the  the number of available licenses and link capacity. 
The success probability approaches one  when both the 
link capacity and the number of licenses are increased.   Note that unilaterally   over-provisioning the link capacity or the number of licenses is not sufficient in order to achieve high values of $s_c$.

The optimization problem described in Section~\ref{sub:optimization} admits a graphical   solution. We start considering the centralized architecture.   Recall that $C'$ is the marginal capacity added to a link of initial capacity $C_0$ resulting in a link with  capacity $C=C'+C_0$.  Varying the values of $C'$ and $L$, we affect the cost given by~\eqref{cost1}.   For each value of the cost variable, $c$, we have a corresponding line in the $(L,C)$ plane characterized by $c = \alpha L + \beta n (C-C_0)$.  Given a set of  cost values, we characterize a family of parallel lines in the $(L,C)$ plane, where $L=1, 2, \ldots$ and $C \ge C_0$.
The minimum value of $c$ for which the corresponding line    intersects the  curve associated to the constraint $s_c \ge s$ corresponds to the optimal centralized solution $c_c^\star$. In the distributed architecture, let $L^{\star}$ be the minimum value of $L$ for which the constraint  $s_d \ge s$ is satisfied.  As in the distributed architecture we assume that licenses and additional state resources are accessed locally, we have  $C'=0$. Therefore,   the optimal distributed solution $c_d^\star$ occurs at  $(L,C)=(L^{\star}, C_0)$.      We compare the centralized and distributed solutions and select the one with minimum cost.

\begin{figure}[h!]
\begin{center}
\bgroup
\hspace{-0.1in}
\setlength\tabcolsep{0pt}  \begin{tabular}{cc} \setlength\tabcolsep{0pt} 
      \includegraphics[width=0.5\textwidth]{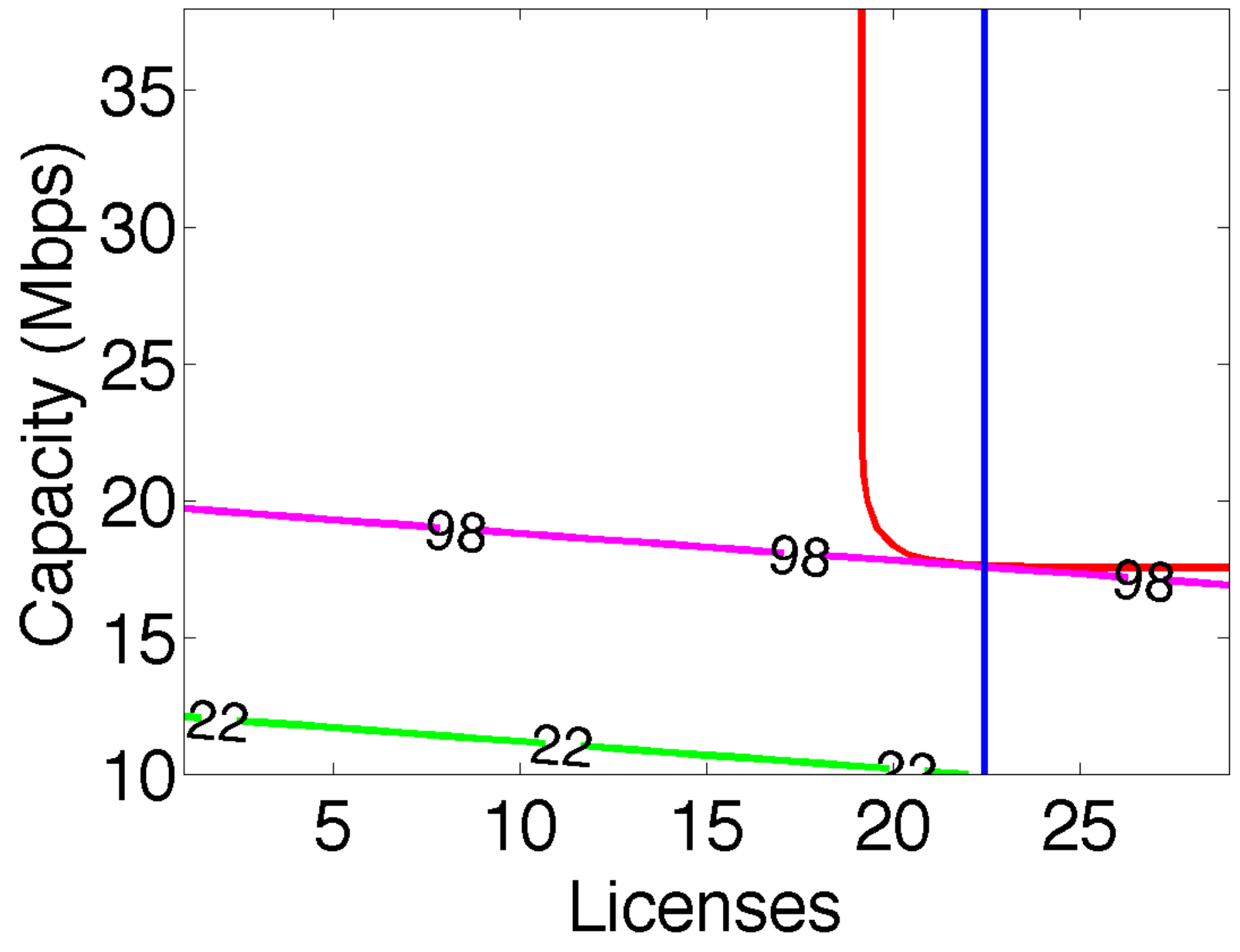} & \hspace{-0.05in}
	  \includegraphics[width=0.5\textwidth]{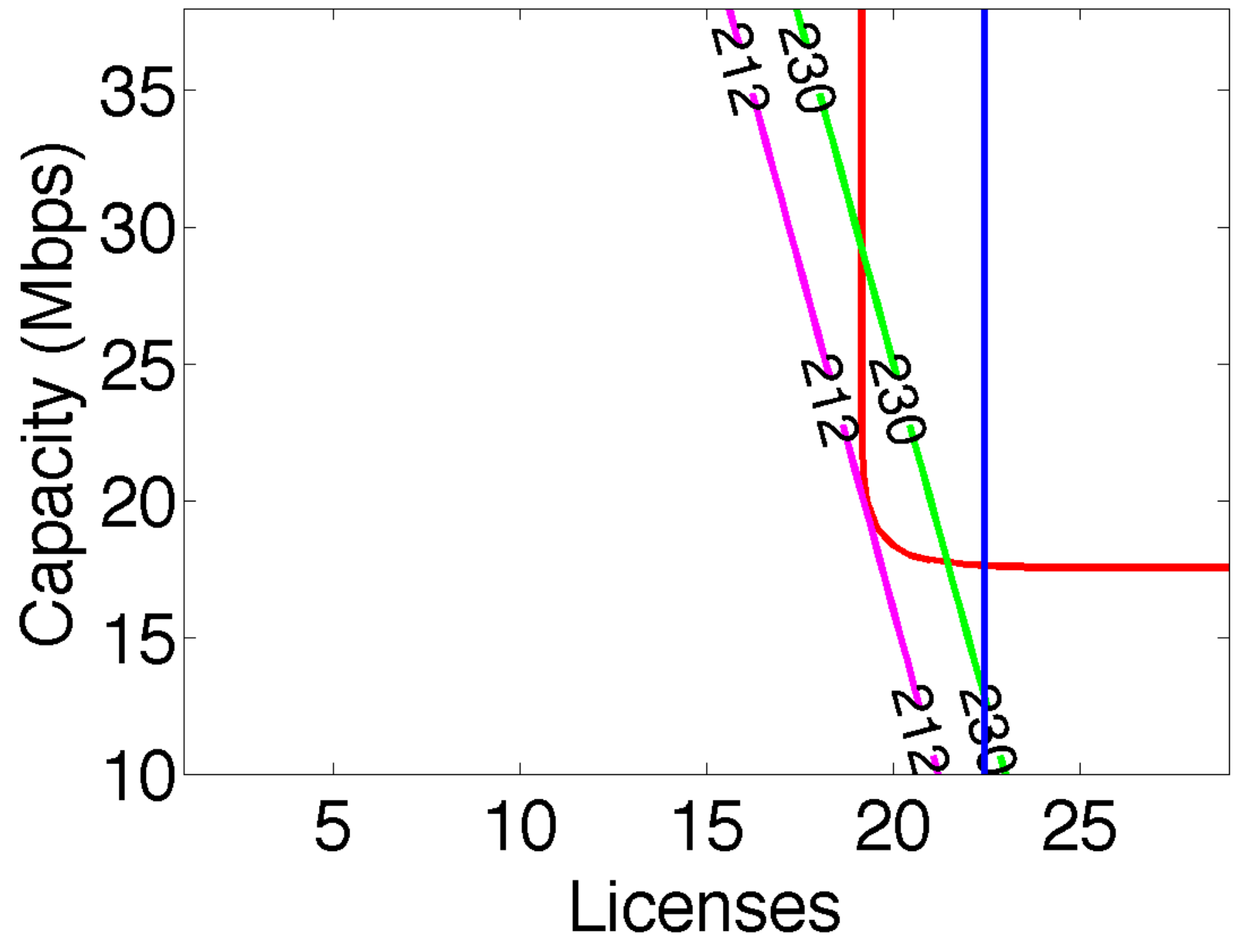} \\
	  (a)  $c_c=L+10C'$  & (b)  $c_c=10L+2C'$  \\
\end{tabular}
\egroup
\end{center}
\vspace{-0.1in}
\caption{Graphical views of optimization problem: (a) $c_c=L+10C'$,  (b) $c_c=10L+2C'$.    ${1}/{\mu} = {1}/{\lambda}= 8$  hours, $S = 30$ users, 
${1}/{\mu} = 8$  hours, ${1}/{r} = 120$ s, $\tau = 0.01$ s, ${1}/{M}=1250 $ Bytes, $\Lambda=900$, $C_0=10$ Mbps.  ``Capacity'' refers to $C=C_0+C'$. }
\label{fig:Optimizations2}
\end{figure}

Figures \ref{fig:Optimizations1}(b),  \ref{fig:Optimizations2}(a) and \ref{fig:Optimizations2}(b) 
illustrate the  graphical solution. 
In all cases the population consists of 30 users, $C_0 = 10$ Mbps and $s=0.95$. When considering distributed solutions, the population is  split  equally between two distinct 
sites.   In 
Figure \ref{fig:Optimizations1}(b) the cost per license is twice the cost per Mbps, meaning that $L+2C'=c$. 
The cost curve marked with ``23'' (magenta line) represents a scenario in which $c=23$, i.e., $L+2C'= 23$. The cost curve marked with ``29'' (green line) represents a scenario in which  $c=29$, i.e., 
$L+2C'= 29$. The red curve is a contour plot of the centralized architecture constraint  wherein $s_c=0.95$, and the blue curve is a contour plot of the distributed  architecture constraint  wherein $s_d=0.95$.
The  intersection  between a cost line and a constraint curve  
corresponding  to the smallest feasible cost  
occurs at the bottom of Figure~\ref{fig:Optimizations1}(b).  
This means that the distributed architecture is the best choice, $c^{\star}=c_d^{\star}=23$, and $L=23$ is the minimum number of licenses that satisfies the SLA requirement. 
To achieve the same SLA, the cost of the centralized architecture would be 29.

Figure \ref{fig:Optimizations2}(a) shows  the graphical solution of the optimization problem when the cost per Megabit/s is five times the cost per 
license. In this case the distributed architecture is  again the best solution.  
The intersection  between a cost line and a constraint curve  
corresponding  to the smallest feasible cost  
occurs at the bottom of Figure~\ref{fig:Optimizations2}(a).  At point $(L,C)=(22,10)$ (or, equivalently, $(L,C')=(22,0)$),  the cost line  $L+10C'=c_d^{\star}$ (magenta line), where $c^{\star}=c_d^{\star}=22$, intersects  the distributed architecture constraint given by $s_d=0.95$ (blue line).  
In this case, to satisfy the service level  agreement the cost of the centralized architecture would be   $c_c^{\star}=98$. The communication cost makes the centralized architecture not viable.

Figure \ref{fig:Optimizations2}(b) shows an example where the centralized architecture outperforms the distributed one.  The cost per license is five times the cost per Megabit/s. In this case, $10L+2C'=c_d^{\star}$ (magenta line), where $c^{\star}= c_d^{\star}=212$, intersects  the red curve associated to the centralized architecture constraint, $s_c=0.95$.  The blue curve corresponding  to the distributed architecture constraint  does not intersect  lines associated to costs smaller than or equal to 230.

As our numerical examples indicate, the proposed methodology allows  an efficient,  principled and graphical exploration of different parameter combinations.  The insights obtained from the model   provide a ``bird's-eye view''  perspective of  different available options.   After a set of  options is selected, detailed and computational intensive  simulations are used to choose the preferred one.

\section{Discussion} \label{sec:discu}

The proposed methodology is a first step in the study of tradeoffs between statistical multiplexing and infrastructure costs in cloud systems.  The analytical model introduced in this paper can be used to investigate the advantages and disadvantages of centralizing signaling servers.  It  allows for what-if analysis of different system parameters, and can be used to explore the state space in a principled way.  
In addition, it can also be used to assist practitioners in setting the state renewal timeout  parameter ($\tau$).  Small values of $\tau$ will yield frequent premature timeouts whereas  larger values  will delay the release of licenses of applications that  unnecessarily remain  active after a user leaves its desktop or after a crash.

We note that in order to obtain a tractable model, we made some simplifying assumptions some of which are discussed below.

\textbf{Channel model characteristics: } the channel model is one of the building blocks of our framework.  In this paper, we consider an M/M/1 queue to model the channel characteristics.  This model can be easily adjusted and adapted according to the needs, while still maintaining  the general framework.

\textbf{Network protocol influence: } we do not model specifics of network protocols such as retransmissions or packet prioritization.  Instead, we take a simplifying approach according to which a timeout occurs if a renewal packet experiences   delay larger than the threshold $\tau$.  The adjustment of the delay-related metrics according to different system characteristics is subject for future work.

\textbf{Available network infrastructure: } we assume an enterprise scenario in which publicly available cloud infrastructures cannot be used due to privacy and safety issues.   Therefore, the infrastructure must be provisioned and planned by a single authority, which motivates the tradeoffs between infrastructure costs and multiplexing benefits discussed in this paper.

\section{Conclusions}
\label{sec:conclusions}

Cloud services are increasingly deployed across multiple and geographically distant sites creating a demand for a holistic performance evaluation before planning 
a converged computing-networking cloud service. We analyzed a case study inspired by a real-world oil and gas industry where a floating license service 
can be distributed among multiple data centers or centralized in a single pool. The centralized case is an example of network-computing cloud service in which
the statistical multiplexing advantages of centralization can be overcome by the corresponding increase in communication infrastructure costs.

We derived an analytical model to evaluate tradeoffs in terms of application requirements, usage patterns and communication costs. The numerical results 
showed that the best solution depends 
on the relation of these several parameters. 
We believe that this model can serve as a guideline for capacity planning of computing and networks resources of floating 
licensing applications and can be a starting point for bridging the computing and networks aspects in an integrated manner.

{
\bibliographystyle{IEEEtran}
\bibliography{eduardo_v8}
}

\end{document}